\documentclass[fleqn,10pt]{wlscirep}
\usepackage[utf8]{inputenc}

\usepackage[T1]{fontenc}
\usepackage{amsmath}
\usepackage{graphicx}
\usepackage{amssymb}
\usepackage{xcolor}
\usepackage{multirow}
\usepackage{todonotes}
\usepackage{subfigure}
\usepackage{dcolumn}
\usepackage{verbatim}
\usepackage{epsf}
\usepackage{hyperref}

\title{Hybrid quantum-classical computation for automatic guided vehicles scheduling}

\author[a]{Tomasz \'{S}mierzchalski}

\author[e,f]{Jakub Pawłowski}

\author[f]{Artur Przybysz}

\author[a]{\L{}ukasz Pawela}

\author[a]{Zbigniew Pucha\l{}a}

\author[b]{M\'aty\'as Koniorczyk}

\author[a]{Bart\l{}omiej Gardas}

\author[c,d]{Sebastian Deffner}

\author[a *]{Krzysztof Domino}

\affil[a]{Institute of Theoretical and Applied Informatics, Polish Academy of Sciences, Ba\l tycka 5, Gliwice, 44-100, Poland}

\affil[b]{HUN-REN Wigner Research Centre for Physics, Konkoly-Thege Mikl\'os \'ut 29-33., Budapest, 1121, Hungary}

\affil[c]{Department of Physics, University of Maryland, Baltimore County, Baltimore, Maryland 21250, USA}

\affil[d]{National Quantum Laboratory, College Park, MD 20740, USA}

\affil[e]{Institute of Theoretical Physics, Faculty of Fundamental Problems of Technology,
Wroclaw University of Science and Technology, 50-370 Wroclaw, Poland}

\affil[f]{Quantumz.io Sp. z o.o., Puławska 12/3, 02-566 Warsaw}

\affil[*]{kdomino@iitis.pl}

\date{July 29, 2023.}


\begin{abstract}

Motivated by recent efforts to develop quantum computing for practical, industrial-scale challenges, we demonstrate the effectiveness of state-of-the-art hybrid (not necessarily quantum) solvers in addressing the business-centric optimization problem of scheduling Automatic Guided Vehicles (AGVs). Some solvers can already leverage noisy intermediate-scale quantum (NISQ) devices. In our study, we utilize D-Wave hybrid solvers that implement classical heuristics with potential assistance from a quantum processing unit. This hybrid methodology performs comparably to existing classical solvers. However, due to the proprietary nature of the software, the precise contribution of quantum computation remains unclear.

Our analysis focuses on a practical, business-oriented scenario: scheduling AGVs within a factory constrained by limited space, simulating a realistic production setting. Our approach maps a realistic AGVs problem onto one reminiscent of railway scheduling and demonstrates that the AGVs problem is better suited to quantum computing than its railway counterpart, the latter being denser in terms of the average number of constraints per variable. The main idea here is to highlight the potential usefulness of a hybrid approach for handling AGVs scheduling problems of practical sizes. We show that a scenario involving up to $21$ AGVs, significant due to possible deadlocks, can be efficiently addressed by a hybrid solver in seconds.
\end{abstract}

\begin{document}

\flushbottom
\maketitle
\thispagestyle{empty}

\section{Introduction}

Quantum computing, a relatively young science field, has developed rapidly in recent years. It is commonly accepted that this novel technology could shape the future. 
Certain tasks are not tractable with a classical computer; quantum computing approaches are good candidates to handle these. For example, Shor's algorithm\cite{shor1994algorithms} proposes a polynomial-time quantum algorithm for the integer factoring problem. 
At its current development level, the improved efficiency offered by quantum computing is restricted to certain computational problems~\cite{1203.5813}. 
Nevertheless, it is widely recognized that we are in the era of noisy intermediate-scale quantum (NISQ) devices~\cite{preskill2018quantum,Brooks_2019}: a few hundred qubits are technologically available, but noise and imperfections still have a relevant impact on their operation. Despite that, the first examples of quantum computing utility have already been presented\cite{arute2019quantum} for some specific physical systems, and there is an increasing research interest in potential industrial applications that are considered\cite{domino2021quantum, PhysRevApplied.14.034009, geitz2022solving}. Currently, the community of quantum computing scientists is searching for practical industrial-scale use cases of industrial problems that can be successfully solved using the available quantum devices~\cite{katzgraber2024searching}. Our intention is to address this quest by tackling the Automated Guided Vehicles (AGVs) scheduling problem. The size and characteristics of the scheduling problem addressed in this paper make it a good candidate to explore noisy-intermediate scale quantum (NISQ)~\cite{preskill2018quantum} technologies in industrial applications.

As current quantum devices are small and noisy, hybrid quantum-classical solvers are being developed to handle optimization problems of sizes that are relevant from the practical application point of view. In these hybrid solvers, a part of the computation is performed on a quantum device: the quantum processing unit (QPU), while the so-called master problem is handled by the CPU (central processing unit). Examples of such an approach are the hybrid solvers developed by the D-Wave company~\cite{dwavehybrid2022}.
In the present paper, we compare the performance of one of these solvers with a state-of-the-art (classical) integer linear programming (ILP) solver when applied for the scheduling of AGVs.  Our findings indicate that in the context of this application, the chosen hybrid solver is comparable with and has the potential to outperform the classical one.

In the broad picture of production scheduling, an approach considered as promising is \emph{Smart Scheduling}\cite{rossit2019industry}. It combines the cyber-physical production systems (CPPS) with the decision support system (DSS). The AGVs scheduling optimization can be treated either as part of the larger industrial system in light of the philosophy of industry 4.0~\cite{stock2016opportunities} (e.g. together with industrial job scheduling) or just as a sole component. For the sake of demonstration, we focus on the latter. However, our model can be re-integrated into a general DSS system. 

Many industrial processes can be optimized, including the design of manufacturing systems and processes, assembly line processes, etc. Within these, we address AGVs scheduling, in particular, timetabling in a short time horizon. The decisions have to be made almost in real-time, requiring fast computational heuristics. The concept of industry 4.0~\cite{luo2018novel} implies that AGVs scheduling must be tied to particular factory characteristics.
To tie our research directly to current business needs, we address a problem that arises in a production environment, i.e. the practice of an actual operating factory. (Its identity and further details are confidential.) We model this particular environment, its configuration, and requirements, addressing the particular needs of the given factory.
In this factory, there is a well-defined space where AGVs can maneuver. They are restricted to moving on dedicated "roads" (uni- or bidirectional) to reach ports, loading places, charging stations, etc. 
Here, the AGVs are controlled by a central system where their scheduling occurs.

AGVs scheduling algorithms are divided into rostering~\footnote{also termed as ``scheduling'' in some works} and routing\cite{zhong2020multi, qiu2002scheduling}. The aim of rostering is to dispatch a set of AGVs to perform certain jobs of equipment transportation within the factory. Then, routing (path planning) aims at finding suitable paths for the AGVs, together with the ''timetable'' and a possible ordering of AGVs passing certain congested places.
These problems are commonly solved via linear programming optimization~\cite{qiu2002scheduling, vivaldini2015comprehensive}. 

The work of Tuan~\cite{le2005intelligent} provides a fair overview of the optimization of AGVs scheduling. Vehicle rostering is discussed both as an offline scheduling problem (where all transportation requests are known in advance) or as online scheduling (where environments are stochastic, and requirements appear on the fly, c.f. the work of\cite{sabuncuoglu2000analysis}). We follow the approach of deterministic offline scheduling, but each time the circumstances change, we recalculate schedules.  
An important constraint (tied to time constraints in scheduling theory) is the maximal window of time 
in which an AGV has to perform a task. The objective is a linear function of variables reflecting completion time costs.
To solve both static and dynamic problems, most authors apply ILP together with a (custom) column generation approach and various methods of optimization. As we consider routes and tasks of AGVs predefined for our optimization problem, column generation is not applicable.

The literature on bigger-scale AGVs scheduling problems has been developed in recent years to catch up with the increasing size of problems in industry practice. For instance, in a paper from 2019 \cite{li2019tasks},
a system with $9$ AGVs had been considered as a large-scale one, while in a more recent paper \cite{kim2023heuristic} from 2023 classical heuristics for much larger AGVs problem were introduced. (Therein, the largest problem was of a few hundred AGVs, but with a computational time of days. The problem solved in a reasonable time of seconds concerns 30-40 AGVs with a particular problem layout.) In
our use case, we introduce a model tailored to the specific needs of
the real-life-driven operational setting, employing proprietary hybrid
quantum-classical solvers for AGVs problems of sizes similar to the above-mentioned. In particular, we
concentrate on certain components of AGVs scheduling, AGVs timing
, and ordering, stressing the problem of deadlock resolution.
As for the actual problem size that a quantum approach can handle, scheduling 2 to 8 AGVs for 10 - 11 tasks (together with routing) on an optical coherent Ising machine was recently reported by Tang et al.\cite{tang2024quantum} (during the preparation of our present contribution). The experimental results therein show that the optical quantum computer has an advantage in computational time over the traditional state-of-art approach.

When compared to the most similar contributions in the literature, our approach is complementary to the work of Geitz et al.~\cite{geitz2022solving}, where the job scheduling in a factory is optimized on a quantum annealer or classical device. 
There, however, the focus is on the factory machines. 
In contrast, we focus on the AGVs traffic, assuming job assignments to industrial machines are fixed. 
Haba et al.\cite{haba2022travel} also addressed a problem of AGVs scheduling with quantum methods. There, however, quantum annealing was applied to the routing of AGVs, as opposed to our problem, which addresses the timing and ordering of the AGVs. 
 
The planning of AGVs paths and task assignment, which is widely
analyzed along with scheduling in literature~\cite{zhong2020multi}, is
not part of our optimization task.  We are required to address AGVs
scheduling in itself, which has less coverage in the literature. 
Our contribution is the first one in this research direction, which is
dedicated to the particular business-driven problem of AGVs scheduling
in a factory with limited space. 
In detail, we introduce a new model for AGVs scheduling (ordering), which is simple enough to be handled by a hybrid quantum-classical approach in a competitive manner and is complex enough to have practical significance, as it is derived from a practical use case. Our high-level goal is to pave the way for the development of new open-source hybrid solvers by the application of the most promising, though proprietary, closed-source D-Wave hybrid solver.
There have been proposals to develop hybrid solvers for scheduling optimization problems\cite{tran2016hybrid,feld2019hybrid}. In these contributions, the optimization task is formulated as a Quadratic Unconstrained Binary Optimization (QUBO) problem, and then its relaxed versions or some subproblems are solved using a quantum annealer, while the remaining part is solved using classical heuristics. However, the transformation of a linear problem into the quadratic representation of QUBO creates a large overhead and requires tuning QUBO parameters. In the present contribution, we use a hybrid solver that inputs the native linear programming model and changes only smaller sub-problems into the QUBO if necessary, which we expect to be more efficient for problems that have suitable linear models, such as AGVs scheduling.

The particular optimization problem handled in our paper appears to be complex but it still
tractable for the meaningful number of AGVs ($21$ AGVs in particular).
The anticipated evolution of Industry 4.0, particularly with respect to AGVs scheduling, is expected to involve more complex scenarios, such as managing 40 or more AGVs within the confined spaces of a factory. Addressing such challenges in a timely manner could become problematic using common heuristics. Hence, new heuristics or computational paradigms may be necessary for such future applications.

This paper is organized as follows. In Section~\ref{sec::method}, we discuss Quantum Annealing. In Section~\ref{sec::AGV_scheduling}, we define our problem and formulate it mathematically. In Section~\ref{sec::calc}, we discuss computational results. In Section~\ref{sec::conclusions}, we draw conclusions, while  Appendix~\ref{sec::appendix_details} is devoted to details on the particular problem of AGVs scheduling.

\section{Quantum annealing}\label{sec::method}

%A recent theoretical study \cite{pirnay2023superpolynomial} points toward possible speedup for approximately solving instances of NP-hard combinatorial optimization problems as long as they are implemented using a fault-tolerant quantum computer.  Inspired by this, we hope that a quantum-ready model can potentially scale better than classical computer algorithms. Hence, developing quantum hardware may open the possibility of efficiently dealing with large problem instances. 
%In particular, we will use quantum annealing, which approximately implements an adiabatic model of quantum computation~\cite{kadowaki1998quantum, farhi2000quantum}. 

Quantum Annealing \cite{apolloni1989quantum} is a heuristic algorithm that operates within the principles of Adiabatic Quantum Computing \cite{kadowaki1998quantum,farhi2000quantum}, particularly for solving optimization problems. In this regime, a given optimization task is encoded into a physical system described by a problem Hamiltonian $H_P$, so that the system's lowest energy state (ground state) corresponds to the solution of the original problem. Initially, the system is prepared in the ground state of another related physical system, described by the initial Hamiltonian $H_0$. Then, the system slowly evolved into the target system $H_P$ by tuning the parameters of the instantaneous Hamiltonian to turn $H_0$ into $H_P$ in a long enough time $T$. According to the adiabatic theorem~\cite{Tanaka2017}, if certain conditions hold, the system should remain in the ground state during the evolution and thus reach the ground state of $H_P$ at the end, yielding a solution to the original problem. Mathematically, the evolution of the system is described by the following time-dependent Hamiltonian:
\begin{equation}\label{eq:evolution}
H(t)=\left(1-\frac{t}{T}\right) H_{0}+\frac{t}{T} H_{P}.
\end{equation}
The quantum annealers which are provided by D-Wave and are applied in our research implement a problem Hamiltonian whose energy is expressed using a 2-local Ising model \cite{lucas2014ising}:
\begin{equation}
    H_P =  \sum_{\langle i, i'\rangle \in 
	E} J'_{i, i'} \sigma^z_{i} \sigma^z_{i'} + \sum_{i} h'_{i} \sigma^z_{i},
 \label{eq::HP_sigma}
\end{equation}
where $\sigma^z$ is a Pauli-Z operator acting on the $i$-th qubit, and $J'_{i, i'}$ and $h'_{i}$ are real values corresponding to pairwise couplings and the external magnetic field, respectively, and $E$ is the graph of the processor topology. The initial Hamiltonian $H_0$ is chosen to consist of a transverse magnetic field,
    $H_0 =  h_{0}\sum_{i} \sigma^x_{i}$
where $\sigma^x_{i}$ is the Pauli-X operator acting on the $i$-th qubit.

Finally, the results of quantum annealing are supposed to minimize the classical Ising problem in Eq.~\eqref{eq::ising}
\begin{equation}\label{eq::ising}
    \min_{(s_1, \ldots s_n) \in \{-1, 1\}^N} \sum_{(i,i') \in E} s_i J'_{i,i'} s_{i'} + \sum_{i \in V} s_i h'_i
\end{equation}
where $s_i$ are spin variables $s_i \in \{-1, 1\}$,  $J'_{i,i'}$ are couplings between spins, and $h'_i$ are local fields. Such a problem can also be easily encoded in
Quadratic Unconstrained Binary Optimization (QUBO) problem, namely:
\begin{equation}\label{eq::qubogen}
   \min_{(x_1, \ldots x_n) \in \{0, 1\}^N} \sum_{(i,i') \in E} x_i J_{i,i'} x_{i'} + \sum_{i \in V} x_i h_i
\end{equation}
where $x_i \in \{0,1\}$ are binary variables, and $s_i = 2 x_i -1$.

In practice, many optimization problems, including our scheduling case, can be modelled as an integer linear program (ILP). The integer variables can be encoded into suitable binaries in various ways~\cite{karimi2019practical,Tamura_2021}. The constraints of the integer programs are taken into account with penalties~\cite{Glover_2022}. The right choice of penalties is a nontrivial problem~\cite{Karimi_2017,gusmeroli2022expedis,quintero2023polyhedral,garcia2022exact} itself.
Although the size of the current quantum annealer is limited in terms of the number of qubits (in our case, approximately $5600$ qubits and $40$ thousands couplings between them organized into the Pegasus topology of QPU~\cite{dwave_machine}), larger optimization problems, that do not fit quantum devices, can be solved via hybrid quantum-classical approach. 
D-Wave offers two hybrid classical-quantum solvers as a part of its service: BQM and CQM~\cite{dwavehybrid2022,dwaveCQM}.

%Currently, studies that address real-life practical problems have started to sporadically
%appear in public domain research.
%Quantum annealing on the D-Wave machine can be performed only on small problems (up to $36$ linear variables or $268$ qubits and $2644$ quadratic couplings) because the size of the annealer (number of qubits and couplings) is limited. Furthermore, the obtained solutions are not feasible due to the annealer's imperfections.
%This observation motivated us to apply hybrid quantum-classical solvers.

The BQM solver inputs QUBO problems and solves them with a portfolio of
classical heuristics.  In the course of the process, certain
subproblems are sent to the quantum processor. In the case of the CQM
solver, the input is a constrained quadratic program, which can be an
ILP.  Its transformation to QUBO is performed within the solver.  The
solver also handles constraints with penalties 
internally.

These solvers are proprietary and closed source; their details are
hidden from the users. Nevertheless, their operation can be understood
according to D-Wave's white paper~\cite{dwavehybrid2022}.  In
particular, the data flow is described as follows.  The solver reads
the input problem.  Then, it invokes a portfolio of heuristic solvers
that run in parallel using classical CPUs and GPUs in a cloud
computing environment.  These heuristics contain a Quantum Module
(QM), which can send queries to QPU.  The quantum results help
classical heuristic search to improve the quality of a current pool of
solutions. After post-processing (removing duplicates, etc.), the
results are forwarded to the user.

The hybrid solvers open the possibility of easily dealing with
bigger-scale problems. For instance, CQM has recently been applied to
the practical rescheduling problems in heterogeneous urban railway
networks~\cite{koniorczyk2023solving}.

%Especially, the use of quantum annealers and hybrid solvers (such as D-Wave hybrid solver ~\cite{dwavehybrid2022}) can handle larger problems than the quantum chip alone. One such hybrid solver has recently been applied to the practical rescheduling problems in heterogeneous urban railway networks~\cite{koniorczyk2023solving}. However, no clear advantage of the hybrid solver over the CPLEX has been presented there. The particular AGVs scheduling problem we analyze in this paper is more demanding even for a moderate problem's size.

\section{AGVs scheduling}\label{sec::AGV_scheduling}

The concept of industry 4.0~\cite{luo2018novel} implies that AGVs scheduling must be tied to the particular factory specifics.
To tie our research directly to current business needs, we address a problem that arises in a production environment, i.e. the practice of an actual operating factory. (Its identity and further details are confidential.) We model this particular environment, its configuration, and requirements, addressing the particular needs of the given factory.
In this factory, there is a well-defined space where AGVs can maneuver. They are restricted to moving on dedicated "roads" (uni- or bidirectional) to reach ports, loading places, charging stations, etc. 
Here, the AGVs are controlled by a central system where their scheduling occurs.

%In the operation of AGVs, \emph{conflicts} may arise as many AGVs compete to use the ``roads''. A conflict is inadmissible when multiple AGVs run on the same segment of the road network, leading to their collision. Thus, it is a basic feasibility requirement against a schedule of AGVs to be conflict-free. The spatial locations where conflicts are possible are called zones~\cite{ho2000dynamic}. Each zone can be occupied by at most one AGVs at a time.

Concerning a factory environment with limited space,
an important issue with AGVs is the possibility of deadlocks, where several AGVs come to such a point that none of them can move forward because the others block each of them.  \cite{le2005intelligent} mentions various deadlock resolution methods:
\begin{enumerate}
    \item Balancing the system workload, i.e. using workload-related dispatching rule~\cite{kim1999agv},
    \item Controlling the traffic at intersections by semaphores~\cite{evers1996automated},
    \item Introducing static or dynamic zones that a limited number of AGVs can occupy for a dynamic-zone strategy for vehicle-collision \cite{ho2000dynamic}
    prevention.
\end{enumerate}
These deadlock resolution strategies can mostly be encoded as
constraints to the ILP.  We have opted for the last approach,
i.e. dynamic zones.
It is worth remembering that deadlock resolution in limited space is challenging, even with a limited problem size (regarding the number of AGVs). 

Our problem can also be understood in the standard metaphor of scheduling theory \cite{pinedo2012scheduling}:  zones are \emph{machines}, and AGVs are \emph{jobs}. Each AGV has to visit a set of infrastructure elements in a given sequence: each job has to be processed by a set of machines in a prescribed order: this is a job-shop \emph{Job Shop} ($Jm$) environment.
Generalizations allowing for having multiple AGVs in a zone would result in a \emph{Flexible Job Shop} ($FJc$) environment; this we will not consider here.
We treat parts of the infrastructure that lie between zones (i.e. where conflicts are not expected) as buffers.
The requirement of a minimal \emph{headways} between AGVs leaving and entering zones and deadlock constraints on bidirectional "roads" (also termed as \emph{lanes}) imply  \emph{blocking} constraints ($block$).
The prescription of the initial availability of AGVs implies \emph{release date} ($r_j$) constraints.
The minimal passing times of AGVs through particular resources are limited: \emph{minimal processing times} ($p_{jm}$) appear. 
In addition, \emph{due dates} ($d_j$) are also prescribed: the AGVs' tasks have to be completed in a given time.
Finally, \emph{permutation} ($prmu$) constraints arise, as AGVs cannot overtake on lanes between zones. Recirculation ($rcrc$), if a given AGV can visit the same zone multiple times during the trip, could also be considered but will not be addressed here as it does not occur in the particular factory we model.

The objective is to ensure that the AGVs finish their travels as soon as possible, considering certain priorities. Hence, the objective is the \emph{travel completion time}, the weighted sum of completion times of each AGV. We shall refer to this as completion time in what follows. With the standard notation of scheduling theory, our problem falls in the class $(J_m |r_j, p_{jm}, d_j, block,  prmu | \sum_j w_j C_j)$.

\begin{figure}[ht]
    \centering
    \includegraphics[width=1.\textwidth]{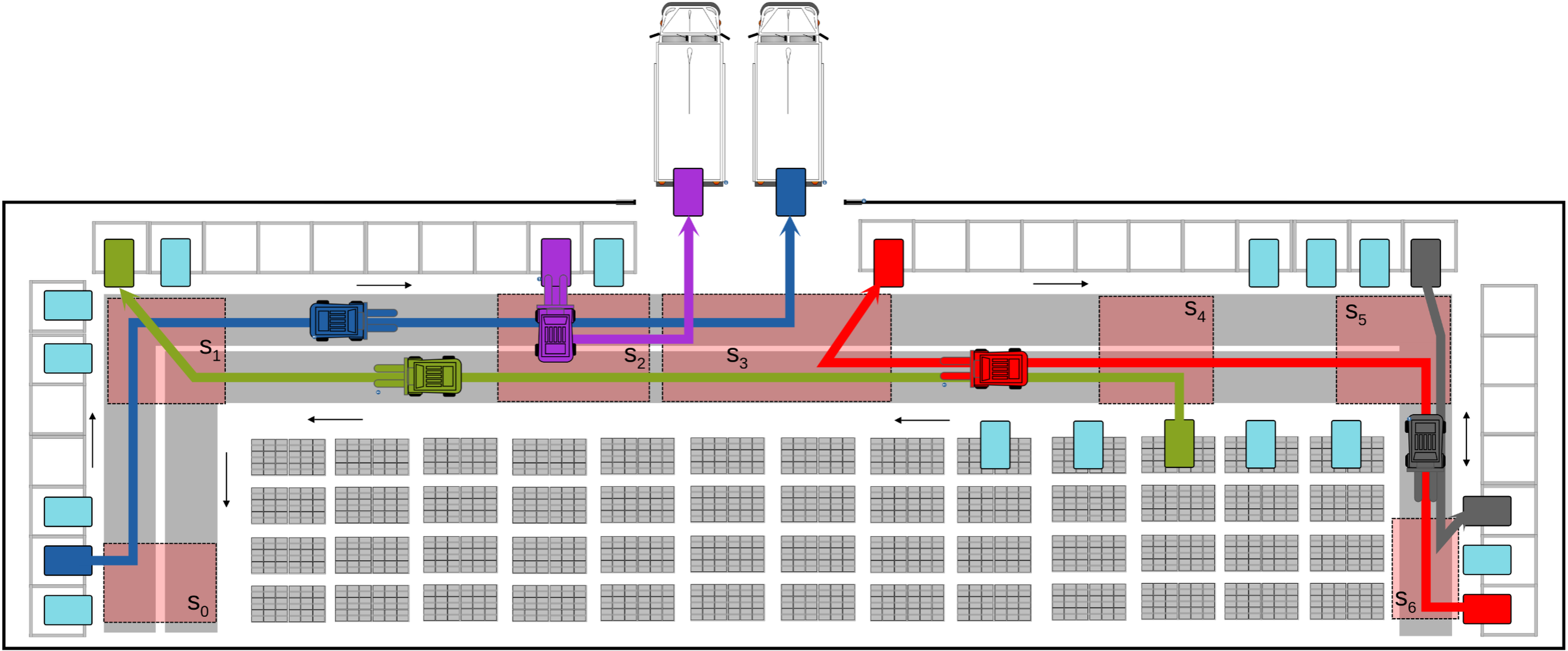}
    \caption{Example of the AGVs scheduling problem given pre-defined AGVs paths as colour lines. First, AGVs are routed to travel between ports (colour rectangles), which is the input to our algorithm. The goal of the algorithm is to timetable AGVs to avoid collisions in zones. In our case zones $s_0, \ldots s_6$ are allegorically assigned spatial areas where AGVs paths intercept start or active ports are located. AGVs paths in terms of zones are {\color{blue} AGV $s_0, s_1, s_2, s_3$}, {\color{black!55!green} AGV $ s_4, s_3, s_2, s_1$}, {\color{red} AGV $ s_6, s_5, s_4, s_3$} {\color{gray} $AGV s_5, s_6$} and {\color{violet} AGV $s_2, s_3$}}.\label{fig:topology_stations}
\end{figure}

The objective is the weighted sum of completion times~\cite{le2005intelligent}. The weights reflect the priorities of AGVs. We assume that AGVs' tasks and paths (represented by the colour lines in Fig.~\ref{fig:topology_stations}) are pre-defined as input for our problem.
Each time these inputs are changed, a new optimization problem is created and recomputed. Such changes in the input can be caused, for example, by disturbances in the AGVs' traffic.

Fig.~\ref{fig:topology_stations} displays an example of the considered topologies. This setup closely resembles the situation in the considered real-life factory (it is distorted to maintain confidentiality).
AGVs paths (color lines) are as follows:
\begin{itemize}
    \item bidirectional lanes (e.g. between $s_5$ and $s_6$ in Fig.~\ref{fig:topology_stations}),
    \item uni-directional lanes,
    \item zones~\cite{ho2000dynamic} the locations where conflicts are possible according to the given (predefined) paths, i.e. where AGVs paths split, join, or where uni-directional lanes start or end.
\end{itemize}
To avoid collisions, we assume that only one AGV can occupy each zone at a given time (This limitation can be lifted in a more elaborate model by increasing the allowed number of AGVs within the zone and introducing some local traffic conditions~\cite{evers1996automated}.) 

\subsection{AGVs scheduling: algorithm}
\label{sec::algorithm}

The practical AGVs scheduling will be performed via the following algorithm triggered at each change of input parameters.
\paragraph{Fixed inputs}
\begin{enumerate}
    \item $J$ - set of AGVs;
    \item $w_j$ - priority weights (as the part of \emph{total weighted completion time} objective);
    \item $d_{\max}$ - the parameter determining the time window, \emph{due time},
    +can be computed from this parameter;
   \item topology of the network as in Fig.~\ref{fig:topology_stations}, containing:
   \begin{itemize}
    \item uni-directional double lanes,
    \item bidirectional single lanes (imposing \emph{blocking} constraints); 
   \end{itemize}
     \item minimal headways between two subsequent AGVs, this is a parameter of \emph{blocking} constraints.
  \end{enumerate}
   \paragraph{Variable inputs}
  \begin{enumerate}
    \item starting points of AGVs in time and space, 
    \item AGVs destinations and paths (from this, the sequence of machines will be determined);
    \item nominal speeds of AGVs, \emph{processing time} constraints can be formulated based on these parameters.
\end{enumerate}
\paragraph{Processing}
\begin{enumerate}
    \item from paths of AGVs create zones where conflicts are possible; see Fig.~\ref{fig:topology_stations},
    \item define AGVs paths by the sequence of zones e.g. $S_j = \{s_{j,1}, s_{j,2} \ldots s_{j,\text{end}-1}, s_{j,\text{end}} \}$, e.g. see caption of Fig.~\ref{fig:topology_stations};
    \item given initial conditions, compute the lowest possible entering and leaving times of each AGV at the zones, assuming there are no collisions between the given AGV and all other AGVs;
    \item encode the problem into ILP: entering and leaving times of AGVs at the zones are integer decision variables, and there are binary order variables determining which AGV leaves a zone first. (A relaxation of the integer constraint on time variables yields a mixed integer program that is also meaningful but not considered here, as quantum devices support integers. In addition, our computational experience shows that the relaxation of the integer constraints on time variables and using a MILP solver does not significantly improve computational time when using CPLEX; the real difficulty is encoded in the order variables.) Lower limits of these variables are determined in point $3$ while upper limits are determined by lower limits and model parameter $d_{\text{max}}$, encode constraints and objective;
    \item solve the problem using a chosen solver (classical, quantum, hybrid, etc.).
\end{enumerate}
\paragraph{Output}
\begin{enumerate}
    \item conflict-free timetable of AGVs encoded as entering and leaving times of AGVs at zones, or correspondingly the order of AGVs therein. 
\end{enumerate}
Steps $1$ - $3$ in \textbf{processing} are pre-processing steps. The higher computational effort is required in step $5$, which deals with optimization. Hence, the analysis of the computation process (both quantum and classical) will refer to step $5$ of the algorithm.
Importantly, the optimization results (values of order variables uniquely defining the order of AGVs at each zone while the traffic beyond zones is conflict-free)  can be directly fed into the application programming interface (API) of the operation control software system of the factory we have modelled. 

\subsection{AGVs scheduling: ILP model}\label{sec::model}

Here, we provide the details of the model in step $4$ of the algorithm, given that step $5$ will be handled in Section~\ref{sec::calc}.

\textbf{Decision variables} include the integer times when AGVs are entering and leaving zones are denoted by  $t_{\text{in / out}}(j,s)$.
As floating point numbers cannot be treated easily, we use integer time variables  $t_{\text{in / out}}(j,s) \in \mathbb{N}$. The time window constraints~\cite{le2005intelligent}, implies that each of these variables fit into the time window of length $d_{\max}$, namely:
\begin{equation}\label{eq::t_vars}
\upsilon_{\text{in / out}}(j,s) \leq t_{\text{in / out}}(j,s) \leq \upsilon_{\text{in / out}}(j,s) + d_{\text{max}}.
\end{equation}
where $\upsilon_{\text{in / out}}(j,s)$ are lower limits computed by assuming the nominal speed of $j$'th AGV and assuming no conflicts (collisions) with other AGVs, according to~\cite{pinedo2012scheduling}. The $d_{\max}$ parameter imposes \emph{due time} constraints, while lower limits are tied to \emph{release date} constraints. 
Let $|S|$ be the number of zones in a particular optimization problem, and 
$|J|$ the number of AGVs therein. 

To determine the order of AGVs at zones, we introduce binary order variables.
\begin{itemize}
\item For zone $s$ passed by AGVs $j$ and $j'$ we have:
\begin{equation}
    y_{\text{in} / \text{out}}(j,j',s) \in \{0,1\}
\end{equation} which is equal to $1$ iff $j$ enters / leaves zone $s$ before $j'$, as the order of AGVs cannot be changed at the zone by assumption; obviously:
\begin{equation}\label{eq::y_one_way} 
    y_{\text{in} / \text{out}}(j,j',s) = 1 - y_{\text{in} / \text{out}}(j',j,s).
\end{equation}
\item For a bidirectional lane (joining zones $s$ and $s'$) used by AGV $j$ heading in one direction and AGV $j'$ heading in the opposite direction, we assign the following order variable:
\begin{equation}
z(j,j',s,s') \in \{0,1\},
\end{equation}
which is equal to $1$ iff $j$ enters such lane (bounded by zone $s$ and $s'$) before $j'$, and zero otherwise, hence,
\begin{equation}\label{eq::y_two_ways}
z(j,j',s,s') = 1 - z(j',j,s',s).
\end{equation}
\end{itemize}

As for the number of variables observe that if an AGV passes a zone, there are two $t$ variables (c.f. Eq.~\eqref{eq::t_vars}); for the AGV entering and leaving the zone. As not all AGVs pass through all zones, we have
 \begin{equation}
 \label{eq::no_t}
     \#t \leq 2 |J| |S|.
 \end{equation}
Concerning order variables, for each pair of AGVs passing a zone there is a pair of these  ($y_{in}$ and $y_{out}$). 
However, the order of AGVs cannot change in a zone (constraints in Eq.~\eqref{eq::ys} will enforce this), a single order variable per zone and AGVs' pair suffices; there are $|J|\left(|J| - 1 \right)/2$ such pairs.
Further, if hypothetically the whole topology consisted of bidirectional lanes, we had $\# z = \# y$. In a more general layout, inequality $\# z \leq \# y$ holds. Hence,
\begin{equation}\label{eq::jlinear}
  \#y \leq \frac{|J|\left(|J| - 1 \right) |S|}{2}, \ \ \
  \#z \leq \frac{|J|\left(|J| - 1 \right) |S|}{2},
\ \ \ \text{and} \ \ \ 
    \# y + \# z + \# t \ll |J|\left(|J| - 1 \right) |S| + 2 |J| |S|.
\end{equation}
For topologies with most uni-directional lanes (as in our example in Section~\ref{sec::algorithm}), we have $\#z \ll \#y$, which leads to the following approximation:
 \begin{equation}\label{eq::no_n}
    \# y + \# z + \# t \ll \frac{|J|\left(|J| - 1 \right) |S|}{2} + 2 |J| |S|.
\end{equation}

\textbf{Objective} is defined  as the weighted sum of completion times \cite{le2005intelligent}, namely:
\begin{equation}\label{eq::objective}
    f = \sum_{j \in J} w_j  \frac{t_{\text{out}}(j,s_{j, \text{end}})  }{d_{\text{max}}}.
\end{equation}
Here $s_{j, \text{end}}$ is the last zone of the path of $j$ AGV, and $w_j$ is the weight tied to the priority of such AGV. (We assume that the AGV's path is conflict-free after leaving the last zone.) This can be referred to as the \emph{total completion time} objective.

\begin{figure}[ht]
\centering
    \includegraphics[width = 0.7\linewidth]{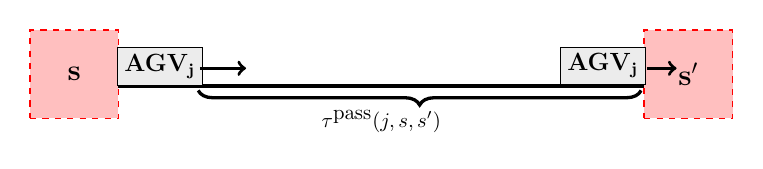}
    \includegraphics[width = 0.7\linewidth]{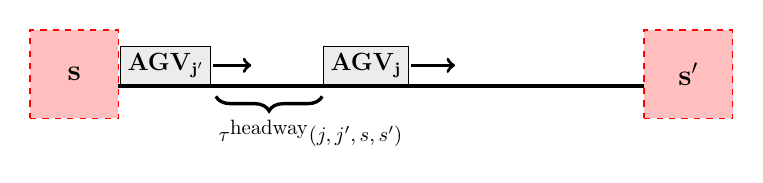}
    \caption{Illustration of \emph{minimal passing time} (upper panel) and \emph{headway} (lower panel) Eq.~\eqref{eq::min_pass_time}}\label{fig::p_headway}
\end{figure}

\textbf{Constraints.} 
The \emph{minimal passing time} constraint (mpt) of AGVs between subsequent zones can be computed from the problem topology and AGVs speeds. For any pair of subsequent zones $(s,s')$ on the route of AGV $j$ we require:
\begin{equation}\label{eq::min_pass_time}  
     t_{\text{in}}(j, s') \geq  t_{\text{out}}(j, s) + \tau^{\text{pass}}(j, s, s'),
\end{equation}
see Fig.~\ref{fig::p_headway} (upper panel). In our model, zones are considered as bottleneck areas. Hence, we allow waiting of AGVs in the buffer before the zone entrance, yielding $\geq$ sign in Eq.~\eqref{eq::min_pass_time}. 

The \emph{minimal headway} constraint (mh) is the model input that determines minimal time spacing between two subsequent AGVs. 
In detail, consider two AGVs $j,j'$ heading in the same direction.
Let $s^*, s, s'$ be the sequence of subsequent zones they both pass. Then: 
\begin{equation}\label{eq::head_out}
    t_{\text{out}}(j', s) + M \cdot y_{out}(j',j,s)   \geq t_{\text{out}}(j, s) + \tau^{\text{headway}}(j, j', s, s'),
\end{equation}
see Fig.~\ref{fig::p_headway} (lower panel), and analogously 
\begin{equation}\label{eq::head_in}
    t_{\text{in}}(j', s) + M \cdot y_{in}(j',j,s)   \geq t_{\text{in}}(j, s) + \tau^{\text{headway}}(j, j', s^*, s). 
\end{equation}
Here, $M$ is a large enough number for inequality to always hold for $y(j',j,s) = 1$ (we used the so-called "big M encoding"). 
If $s$ does not have a successor in the sequence, we do not consider Eq.~\eqref{eq::head_out}. Analogously, if $s$ does not have the predecessor, we do not consider Eq.~\eqref{eq::head_in}. 

The \emph{deadlock} on bidirectional lane constraint (d) is defined as follows.
Let the pair $(s,s')$ be two zones connected by the single bidirectional lane and $j$ and $j'$ be two AGVs first going $s \rightarrow s'$ and the second $s \leftarrow s'$. Then: 
\begin{equation}\label{eq::stl_cond}
     t_{\text{out}}(j', s') + M\cdot z(j',j ,s', s) \geq t_{\text{in}}(j, s')
\end{equation}
and
\begin{equation}
    z(j,j',s,s') = y_{in}(j,j',s) =  y_{in}(j,j',s')
\label{eq::stl_order}
\end{equation}
Equation~\eqref{eq::stl_order} yields that the order of AGVs heading in opposite directions cannot change between zones $s$, $s'$ (including zones itself), as they cannot meet at the bidirectional lane, which would lead to the deadlock.

In the \emph{zone constraint} (zc), we assume that the zone can be occupied only by one AGV at a time.  This is to avoid collisions within the zone. 
Hence, AGV $j'$ can enter the zone after $j$ leaves it (provided $j$ is first on the zone, i.e. $y_{in}(j',j,s) = 0$):
\begin{equation}\label{eq::s_constraint}
    t_{\text{in}}(j', s) + M \cdot y_{in}(j',j,s) \geq t_{\text{out}}(j, s)
\end{equation}
Then we also have to include minimal passing time over the zone $\tau^{\text{zone}}$
\begin{equation}\label{eq::min_stay}
t_{\text{out}}(j, s) \geq  t_{\text{in}}(j, s) +
    \tau^{\text{zone}}(j, s).
\end{equation}
In the scheduling theory language: the assumption that one AGV can occupy the zone only defines the \emph{job shop} machine environment, and the prescription of the passing time over zones imposes \emph{processing time} constraints.

As AGVs are, in general, moving with similar speed, we assume that they cannot overtake both on lanes and zones. This \emph{no overtake} (no) constraint requires the order of such AGVs to be maintained. (This is the \emph{permutation constraint}.)
In detail, let $j$ and $j'$ be a pair of AGVs heading in the same direction, passing the sequence of zones: $s_1, s_2, \ldots s_k$. Then:
\begin{equation}\label{eq::ys_no_mo_line}
    y(j,j', {s_1}_{out}) = y(j,j', {s_2}_{out}) = \ldots = y(j,j', {s_k}_{out})
\end{equation}
(Note that if the route of two AGVs splits and joins later, the condition in Eq.~\eqref{eq::ys_no_mo_line} has to be modified). 

As mentioned before, we also assume that AGVs cannot overtake within a zone:
\begin{equation}\label{eq::ys}
    y_{in}(j,j',s) = y_{out}(j,j',s).
\end{equation}
Let us now estimate the number of constraints.
The \emph{minimal passing time} (mpt) constraint, there is a single inequality as in Eq.~\eqref{eq::min_pass_time} for each AGV and each zone passed by this AGV. Then, taking the upper limit of each AGV passing each zone, we have
\begin{equation}
    \#_{\text{mpt}} \leq |J| |S|.
    \label{eq::mpt}
\end{equation}
The \emph{minimal headway} (mh) constraints are required for each pair of AGVs following each other at each zone. There are roughly $|J|^2/2$ such pairs. To obtain an upper limit we consider all AGVs passing all zones. Then, for each AGV in the pair and each zone, we have one constraint from  Eq.~\eqref{eq::head_out} and one from Eq.~\eqref{eq::head_in};
\begin{equation}
    \#_{\text{mh}} \leq 2 \frac{|J|^2}{2}|S| = |J|^2 |S|.
    \label{eq:mh_constr}
\end{equation}
Analogously , we would expect from Eq.~\eqref{eq::stl_cond} the same limit for the number of \emph{deadlock} (d) constraints;
\begin{equation}
    \#_{\text{mh}} + \#_{\text{d}} \leq 2|J|^2 |S|.
    \label{eq::d_constraint}
\end{equation}
To approximate the number of \emph{zone constraints} (zc), the upper limit is obtained again with the hypothetical assumption that all AGVs pass all zones. Then for each zone and each AGV we have $|J|-1$ inequalities from Eq.~\eqref{eq::s_constraint} and one inequality from Eq.~\eqref{eq::min_stay}, yielding
\begin{equation}
\label{eq:zc_constraint}
    \#_{\text{zc}} \leq  |J|^2|S|.
\end{equation}

Hence, the number of \emph{minimal passing time} (mpt) constraints (see Eq.~\eqref{eq::mpt} ) is linear in $|J|$, while the number of all other constraints (zc, mh, d) is quadratic in $|J|$. Then, for a large enough $|J|$, we can assume:
\begin{equation}\label{eq::no_m}
    m  = \#_{\text{mpt}} + \#_{\text{zc}} + \#_{\text{mh}} + \#_{\text{d}} \approx \#_{\text{zc}} + \#_{\text{mh}} + \#_{\text{d}} \leq 3 |J|^2|S|,
\end{equation}
Finally, the number of \textbf{no overtake} (no) constraints in Eq.~\eqref{eq::ys_no_mo_line}
\begin{equation}\label{eq::mo_limit}
    \#_{\text{no}} \leq \frac{|J|^2|S|}{2}
\end{equation}
is also expected to be linear in $|S|$ and quadratic in $|J|$. 

Altogether, the problem size (in terms of the number of constraints and binary variables) scales linearly in the number of zones $|S|$ and quadratically in the number of AGVs $|J|$. The quadratic scaling can cause difficulties when applying the ILP model to a large number of AGVs. As such, this AGV problem seems to be a good use-case for the quantum approach described at the beginning of this section.

\subsection{QUBO model}\label{sec::bqm}

The most general form of inequalities in our model is in  Eq.~\eqref{eq::head_out} or Eq.~\eqref{eq::head_in}. Based on these, we present a detailed path of transformation of these inequalities into QUBO.
Let us contract $t$-variables and $y$-variables to vectors with elements $t_j$ and $y_i$ (and define the respective vector for the constant terms $c_i$). Then, from the aforementioned equations %(or similar)
, we have:
\begin{equation}\label{eq::constraint_in_slack}
    t_j - t_{j'} - M y_{i}  +\xi_i = - c_{i}.
\end{equation}
To replace inequalities with equalities, we use slack variables
\begin{equation}
    0 \leq \xi_i \leq \bar \xi_i  \ \text{ where } \ 
\bar \xi_i = -\min( t_j - t_{j'} - My_i)- c_i =  - \upsilon_{j} + \upsilon_{j'} + d_{\max} + M_i - c_i = d_{\max} + M_i,
\end{equation}
as $\min(t_j) = v_j$, and $\max(t_j) = v_j + d_{\max}$ from Eq.~\eqref{eq::t_vars}, and 
%we expect
$\upsilon_{j'} = \upsilon_{j} + c_i$. Observe that some of the inequalities such as these of the \textbf{minimal passing time} constraints in Eq.~\eqref{eq::min_pass_time} do not have order variables ($y$ or $z$), for such variables $\bar \xi_i = d_{\max}$. 

To convert the constrained problem in Eq.~\eqref{eq::constraint_in_slack} into unconstrained, we use the penalty method: each constraint violation is penalized by the hard constraint penalty $p>0$. Such a penalty has to be sufficiently large not to be overruled by the objective. Then, the optimization problem has the form:
\begin{equation}
\begin{split}
    \textrm{min.}_{t_1, \ldots, y'_1, \ldots, \xi_1, \ldots} \ \  p \sum_{(j,j',i) \in I}( { t_j - t_{j'} - M_i y'_{i}  +\xi_i + c_{i} )^2 } +  p \sum_{(i,i') \in {I'}} (y'_i - y'_{i'})^2 + \text{objective}.
    \end{split}
    \label{eq::qubo}
\end{equation}
The first sum yields inequalities (mpt, mh, d, zc constraints) and takes $M_i = 0$ if no order variable is included. The set $I$ contains all indices tied to these constraints, yielding
$|I| = m$ (c.f. Eq.~\eqref{eq::no_m}). The number of slack variables is also equal to $m$, as we have one slack variable per inequality. The second sum is for equates to (no) constraint 
and (d) constraint. Then, ${I'}$ contains all indices tied to this constraint, and $|{I'}| =\#_{\text{no}} + \#_{\text{d}}$.

Finally (following Eq. (5) in \cite{karimi2019practical}),  we replace all $t_j$ and $\xi_i$ with the corresponding monomial of bits $\sum_{k}d_k b_k$ where $b_k$ are new bits variables. This transforms the quadratic unconstrained model into the quadratic unconstrained binary model. We have $\# t$ t-variables (with $d_{\max} + 1$ distinct values) and $m$ slack variables (with at most
$d_{\max} + \max_i \xi_i  + 1$ distinct values). Referring to Eq.~\eqref{eq::no_m}, the number of binary variables in QUBO representation can be limited by
\begin{equation}
\label{eq::no_qubo}
\# \text{QUBO} \leq \# t  \big\lceil \ln_2 (d_{\text{max}} + 1)\big\rceil + m \big\lceil \ln_2 (d_{\text{max}} + \max_i \xi_i  + 1)\big\rceil  + \#y + \#z.
\end{equation}
and referring to Eqs.~\eqref{eq::no_t}~\eqref{eq::no_n}~\eqref{eq::no_m}:
\begin{equation}\label{eq::no_qubo_JS}
\# \text{QUBO} \leq 2 |J| |S| \big\lceil \ln_2 (d_{\text{max}} + 1)\big\rceil + 3 |J|^2 |S| \big\lceil \ln_2 (d_{\text{max}} + \max_i \xi_i  + 1)\big\rceil  + \frac{|J|\left(|J| - 1 \right) |S|}{2} = O( |J|^2 |S|).
\end{equation} 

The size of the problem (in terms of the number of QUBO variables) scales linearly with the number of zones and quadratically in the number of AGVs. The number of constraints per variable (the mean vertex degree of the graph model) is tied to the expansion of quadratic terms in Eq.~\eqref{eq::qubo}. In Table.~\ref{tab::ising_size}, the sizes of actual problems in terms of Ising variables (derived directly from QUBO) are presented. Hence, this is the number of the so-called logical qubits. Upon the procedure of using a hardware annealer, the problem has to be embedded into the physical topology of the annealer, which is a nontrivial procedure and is not always possible~\cite{cai2014practical}. The actual problem solved on the hardware is thus expressed in terms of physical qubits, and a logical qubit can possibly be implemented using multiple physical qubits.

Apart from the two smallest problems from Table.~\ref{tab::ising_size}, the others did not embed into the Pegasus chip, see Fig.~\ref{fig::details}. 
For larger instances, the hybrid, quantum-classical approach is necessary. For example, for the $15$ AGVs and $7$ zones problem we have $696$ variables (upper limit according to Eq.~\eqref{eq::no_n} was $945$), and $1378$ linear constraints (upper limit according to Eq.~\eqref{eq::mo_limit}~\eqref{eq::no_m} was $5513$). 

%These limits are quadratic in the number of AGVs and linear in the number of zones. We may expect the problem size to be quadratic with the number of AGVs as the number of zones is constant. If, however, the number of AGVs increases further,  more zones will appear where collisions occur, and therefore, the problem size will scale worse than quadratic.

\begin{table}
    \centering
    \begin{tabular}{|c|c|c|c|c|} \hline 
         $|J|$ /  $|S|$ & \begin{tabular}{c}	n.o. qubits \\ (vertices) \end{tabular}	& \begin{tabular}{c} n.o. quadratic couplings	\\ edges \end{tabular} & edge density~\cite{darlay2012dense} $\frac{edges}{full graph}$ & n.o. linear fields \\ \hline
         2  / 4 &  122& 1066	& 0.14 & 121\\ \hline 
         4  / 4  &  268& 2644& 0.07 & 267\\ \hline 
         6  / 7 &  796& 11954& 0.04 & 782\\ \hline 
         7  / 7 &  1204& 19084& 0.03 & 1183\\ \hline 
         12 / 7 &  6357& 116422& 0.006 & 6250\\ \hline 
         15 / 7 &  7343&  134415& 0.005 & 7219\\ \hline
         21 / 7 &  13892 &  259702 & 0.005 & 13648\\ \hline
    \end{tabular}
    \caption{Sizes of analyzed AGVs problems, in terms of Ising variables. Note, that the edge density is smaller than the one for a corresponding full graph, and it decreases with increasing problem size (up to 3 orders of magnitude for the largest considered problem).}
    \label{tab::ising_size}
\end{table}

\section{Computational results and discussion}\label{sec::calc}

\begin{figure}[ht]
\centering
    \includegraphics[width = 0.8\linewidth]{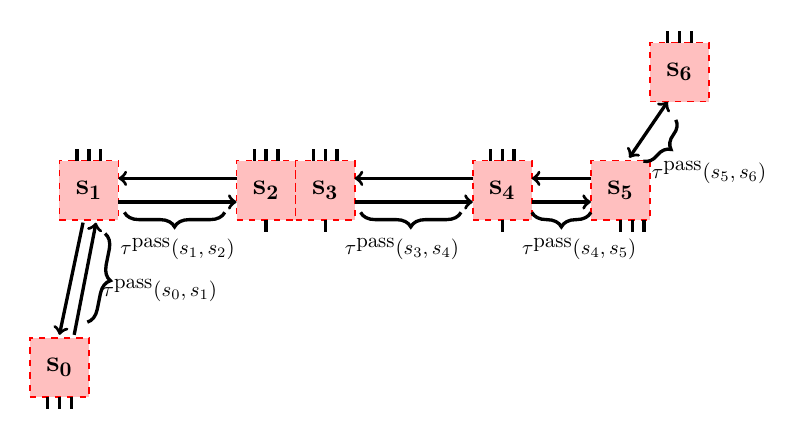}
    \caption{Example network with $7$ zones derived from Fig.~\ref{fig:topology_stations}}\label{fig::network}
\end{figure}

For numerical calculations, we used examples of problems similar to those presented in Fig.~\ref{fig:topology_stations}, but with different numbers of AGVs and zones.
This is a typical setting in an industrial system, as various AGVs are in traffic at various times, and collision zones form dynamically. 
As a typical example of the particular parameter setting with $7$ zones see  Fig.~\ref{fig::network}.
The actual values of parameters can be read from the topology, AGVs' speeds, and zone locations.
The example with $7$ zones and $7$ AGVs will be discussed in more detail in Appendix~\ref{sec::appendix_details}.

From an optimization point of view, one of our goals is to examine the
scaling of the problems' computational time, number of variables
and constraints with the size of the problem in terms of $|J|$ and
$|S|$ (namely the number of AGVs and zones). We start with $2$ AGVs
and $4$ zones and end with $21$ AGVs and $7$ zones. The sizes of the
instances we have solved are tabulated in Table.~\ref{tab:sizes}. It is interesting to compare this to an analogous problem in the railways solved by us in a previous contribution~\cite{koniorczyk2023solving}: while there we had $3$ to $5.5$ constraints per variable, the AGVs problem is significantly sparser (approximately $2$ constraints per variable), making it more likely to be better handled by a quantum device, where the embedding on the device with limited degree connectivity is performed. Note, however, that if the considered AGVs problem address a more branched topology (e.g. with loops, etc.), this would result in a denser problem, more similar to the analogous railway-related one, with more constraints per variable. This would be less suited for the quantum-based approach.

\begin{table}
\centering
\begin{tabular}{|l|lll|lll|}
\hline
\multirow{2}{*}{\begin{tabular}[c]{@{}l@{}} $|J|$ / $|S|$ / $d_{\max}$\end{tabular}} & \multicolumn{3}{l|}{actual ILP} & \multicolumn{3}{l|}{upper limit}  \\ \cline{2-7} 
 & \multicolumn{1}{l|}{\begin{tabular}[c]{@{}l@{}}n.o. vars \\ int / bin\end{tabular}} & \multicolumn{1}{l|}{\begin{tabular}[c]{@{}l@{}}n.o. \\ equalities\end{tabular}} & \begin{tabular}[c]{@{}l@{}}n.o. \\ inequalities\end{tabular} & \multicolumn{1}{l|}{\begin{tabular}[c]{@{}l@{}}n.o. vars \\ Eq.~\eqref{eq::no_n}\end{tabular}} & \multicolumn{1}{l|}{\begin{tabular}[c]{@{}l@{}}\ equalities \\ Eq.~\eqref{eq::mo_limit}\end{tabular}} & \begin{tabular}[c]{@{}l@{}} inequalities \\ Eq.~\eqref{eq::no_m}\end{tabular} \\ \hline
2  / 4 / 10 & \multicolumn{1}{l|}{16 \ 12/4}  & \multicolumn{1}{l|}{4} & 16  & \multicolumn{1}{l|}{20}    & \multicolumn{1}{l|}{ 8}  & 48  \\ \hline
4  / 4 / 10  & \multicolumn{1}{l|}{36 \ 20/16} & \multicolumn{1}{l|}{10}   & 34  & \multicolumn{1}{l|}{56}   & \multicolumn{1}{l|}{ 32}  & 192  \\ \hline
6  / 7 / 40 & \multicolumn{1}{l|}{78 \ 38/40}   & \multicolumn{1}{l|}{32}  & 80 & \multicolumn{1}{l|}{189}   & \multicolumn{1}{l|}{126 }   & 756  \\ \hline
7  / 7 / 40 & \multicolumn{1}{l|}{118 \ 48/70}   & \multicolumn{1}{l|}{55}  & 127 & \multicolumn{1}{l|}{245}   & \multicolumn{1}{l|}{ 172}   &  1029 \\ \hline
12 / 7 / 40 & \multicolumn{1}{l|}{596  \ 102/494}    & \multicolumn{1}{l|}{429} & 766  & \multicolumn{1}{l|}{630}  & \multicolumn{1}{l|}{504 }  & 3024  \\ \hline
15 / 7 / 40 & \multicolumn{1}{l|}{696 \ 118/576}    & \multicolumn{1}{l|}{495} & 883 & \multicolumn{1}{l|}{945}  & \multicolumn{1}{l|}{  788}  &  4725   \\ \hline
21 / 7 / 40 & \multicolumn{1}{l|}{1302 \ 168/1134}    & \multicolumn{1}{l|}{1007} & 1705 & \multicolumn{1}{l|}{1764}  & \multicolumn{1}{l|}{  1544}  &  9261   \\ \hline
\end{tabular}
\caption{Sizes of ILPs of the analyzed AGVs problems. The notion of upper limits is described in Section~\ref{sec::model}. We have approximately $2$ constraints per variable (equality and inequality), which makes the problem sparse in comparison with the railway counter-example in~\cite{koniorczyk2023solving}. }\label{tab:sizes}
\end{table}

\begin{figure}
    \includegraphics[width = 1.0\linewidth]{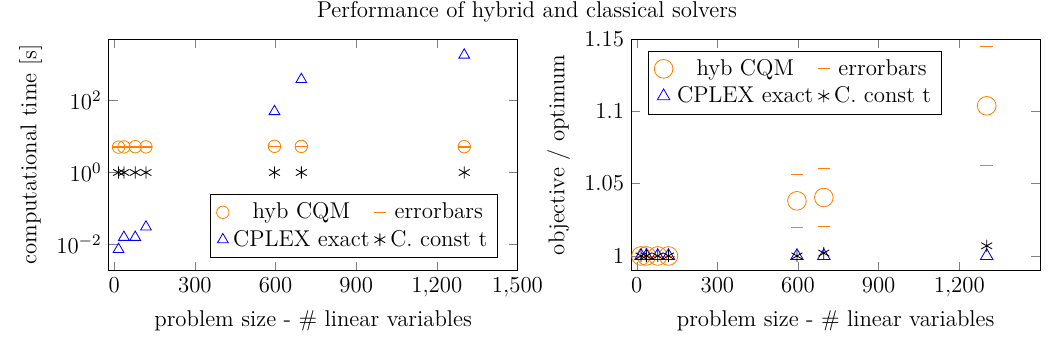}
    \caption{Comparison of performance of Classical CPLEX exact and approximate (achieved by setting constant computational time) with hybrid quantum-classical solver, CQM in particular. All presented solutions were feasible. For CPLEX exact on the largest instance ($1302$ variables) the time limit of $30$ minutes was used, as it was not possible to achieve the certified output in a reasonable time, $6$h time limit CPLEX computation yields the same results. Error bars were computed by performing $10$ independent realizations of experiments of CQM computation and calculating the standard deviation over realizations.} \label{fig::comaprison}
\end{figure}

All classical (that is, not quantum) calculations were performed on the consumer-grade computer described in Table.~\ref{cpu}, and quantum calculations were performed on D-Wave's Advantage\_systems6.2 quantum annealer, whose physical properties are described in~\cite{dwave_machine}. In short, the quantum device has over $5600$ qubits and $40$ a thousand couplings between them, organized into the Pegasus topology of QPU. Concerning pure quantum computing, we were able to fit to real quantum device only the first two smallest problems from Table.~\ref{tab:sizes}. The particular sizes of these problems in terms of Ising variables are presented in Table.~\ref{tab::ising_size}. The problem of $36$ ILP variables and $44$ ILP constraints (equalities and inequalities) was converted into $268$ qubits Ising problem with $2644$ (quadratic) couplings between. However, even for these two small problems, we did not achieve the feasible solution. From this, we can conclude that quantum annealers are too small and too prone to errors, and hence, for more accurate solutions to larger problems, we opt for hybrid quantum-classical solvers. 

\begin{table}
    \centering
    \begin{tabular}{|c|c|}
        \hline
        OS & Windows 10 Pro N\\ \hline
        Type & 64 bits \\ \hline
        CPU & 11th Gen Intel(R) Core(TM) i5-11600K @ 3.90GHz\\ \hline
        RAM & 32 GB\\ \hline
         GPU & Intel(R) UHD Graphics 750 \\ \hline
    \end{tabular}
    \caption{Technical specifications of the classical computer used for computations.}
    \label{cpu}
\end{table}

\begin{figure}
    \includegraphics[width = 1.0\linewidth]{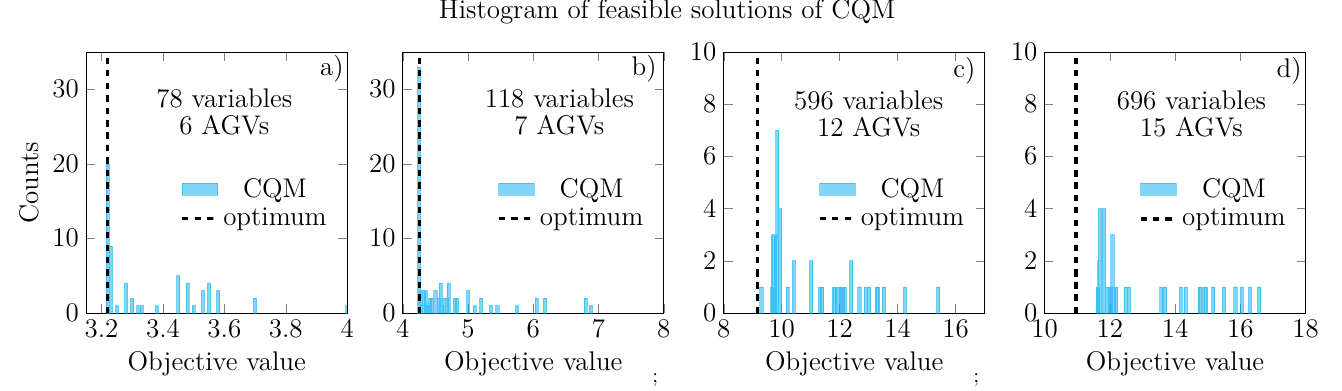}
    \caption{Histograms of objectives of feasible CQM solutions for various problem sizes: $6$ AGVs a), $7$ AGVs b), $12$ AGVs c), $15$ AGVs d). For each run, approximately a hundred samples were returned. Histogram spreads are measured by the standard deviation (std): $6$ AGVs std = $0.19$, $7$ AGVs std = $0.62$, $12$ AGVs std = $1.51$, $15$ AGVs std = $1.9$.}\label{fig::feasibility}
\end{figure}

\begin{figure}
    \includegraphics[width = 1.0\linewidth]{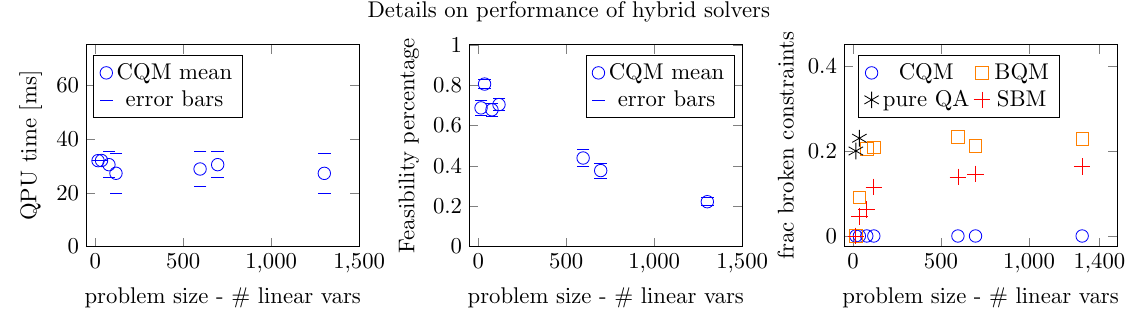}
    \caption{Details of hybrid solvers. The QPU time is in the left panel, the feasibility percentage is in the middle panel, and the number of broken constraints of the solution is in the right panel (pure QA - pure quantum annealing, SBM - simulated bifurcation). Observe that the output of the CQM solver was satisfactory in terms of feasibility for all instances.}\label{fig::details}
\end{figure}

Indeed, the CQM hybrid solver, which is convenient also because it accepts ILPs directly, has always provided some feasible solutions in our problems. We have therefore made a comparison of 3 approaches: the exact CPLEX ILP solver which always solves the problem to optimality, an approximate CPLEX solution, achieved by setting constant computational time for CPLEX solver, and the CQM hybrid solver as an approximate heuristic. Our comparison is summarized in Fig.~\ref{fig::comaprison}.
In terms of computational time, CQM has outperformed the classical exact CPLEX solver for large problems ($596$ and $696$ linear variables). It has to be noted that in each of the CQM computations, a nonzero QPU time was reported (see Fig.~\ref{fig::details}, left panel), indicating that the QPU was active during the calculation. From the reported QPU computational times we can conclude that rather a small amount of computation was performed on the QPU, and the QPU time does not scale with the problem size. The probable explanation of this observation is that the D-Wave hybrid solvers use "Quantum Backstops". When stuck in a local minimum, the solver uses a quantum approach on selected sub-problem to move away from such a local minimum. The number of such "Quantum Backstops", and tied to them the QPU time, may not depend on the problem size.
The largest problem was of $21$ AGVs and $7$ zones, which is the sound size concerning the real live factory environment. As for the quality of the solutions, the CQM yields slightly higher objective value than the exact solution. 
 This overview shows that the CPLEX approximate solver, performing all computations in $1$ second, is fast enough for almost online scheduling and dynamics response of our algorithm. 
Hence, such results would be admissible for practical application in real-life factories. 
In the future, when better quantum annealers will be available, they may become more efficient than classical heuristics. Concerning railway counterpart problem in ~\cite{koniorczyk2023solving} and analogous real live problems of $817$ variables and over $3000$ constraints, the performance of the CQM solver therein could compete only with exact CPLEX in terms of computational time but at larger cost of the solution quality than in the AGVs case.

As the CQM solver returns multiple feasible solutions at a time, in Fig.~\ref{fig::feasibility}, we present the histograms of these for problems of various sizes. Observe that for small problems, the results concentrate around the optimum. In contrast, for large problems, the histogram spreads towards higher-than-optimum objective values,  the standard deviation value presented in caption grows by the order of magnitude, if we compare the $6$ AGVs problem with the $15$ AGVs one.
It is important to note that even for large problems, there are multiple feasible and potentially valuable solutions obtained.
The results with the CQM solver are competitive and illustrate an efficient application of hybrid algorithms. The details of a particular case and its solutions, including a detailed comparison of classical and different quantum solutions, are presented in Appendix~\ref{sec::appendix_details}.

For the largest problem of $21$ AGVs and $7$ zones ($1302$ linear variables), the problem is still tractable using classical heuristics. As we expect the size of the problem to scale quadratically with the number of AGVs, other heuristics may be required to handle larger problems in expanding industrial environments; a hybrid quantum-classical approach may be one of these. This, however, would also require the improvement of the hybrid solvers as the feasibility percentage of CQM decreases with the problem size, as it can be observed in Fig.~\ref{fig::details}, middle panel. If the linear scaling between feasibility percentage and the number of linear variables holds, the biggest problems that can be solvable using the current CQM can have approximately up to $1000$ variables. 

Recall that our first approach was to use the BQM solver. In order to do so, the ILP has to be converted to QUBO. Section~\ref{sec::bqm} describes the details of this transcription. The constraints were taken into account with penalties using ad-hoc penalty coefficients. We have found that except for the smallest problems, the BQM solver does not return high-quality solutions (see Fig.~\ref{fig::details}, right panel). Notably, the solutions were not feasible: a more advanced and systematic determination should replace our ad-hoc choice of the penalty coefficients. The interesting observation here is, however, that for BQM, the percentage of non-feasible solutions is constant for larger instances and equal to these from the pure Quantum Annealing for smaller ones. Hence, here the unfeasibility of Quantum Annealing at the lower scale may have an effect on the BQM performance on the larger scale. For the comprehensive analysis, in Fig.~\ref{fig::details}, the right panel we compare two QUBO base solvers, already mentioned proprietary, closed source BQM and open source Simulated Bifurcation Machine (SBM) solver. The SBM behaves better than the BQM, but still does not return feasible solutions for large instances. This observation paves the way for the development of new quantum-inspired solvers, as the algorithm behind SBM is known.

Drawing inspiration from quantum adiabatic optimization with a network of nonlinear quantum oscillators, SBM exploits the phenomenon of bifurcation in a system of classical, nonlinear system of Hamilton equations\cite{goto2019sbm}:
\begin{align}
    H &= \frac{a_0}{2} \sum_{i=1}^{N} p_i^2 + \sum_{i=1}^{N} \left(\frac{q_i^4}{4} + \frac{a_0 - a(t)}{2} q_i^2 \right) - c_0
    \sum_{i=1}^{N} \left(h'_i q_i + \frac{1}{2}\sum_{j=1}^N J'_{ij} q_i q_j\right) \\
    \dot{q}_i &= \frac{\partial H}{\partial p_i} = a_0 p_i \label{eq:sbm2} \\ 
    \dot{p}_i &= -\frac{\partial H}{\partial q_i} = -\left[q_i^2 + a_0 - a(t)\right] q_i + c_0 \left(\sum_{j=1}^{N} J'_{ij} q_j + h'_i\right) \label{eq:sbm}
\end{align}
It can be interpreted as a motion of particles with mass \(a_0^{-1}\), position \(q_i \in \mathbb{R}\), and momentum \(p_i\in \mathbb{R}\), in a time-dependent potential and coupled via Ising-like interaction. Matrix \(J'\) and vector \(h'\) encode the Ising minimization problem in question, cf. Eq\eqref{eq::ising}. Numbers \(a_0\) and \(c_0\) are constant hyperparameters (in our calculations, we use \(a_0=1\) and \(c_0 = 1/\lambda_{\text{max}}\), where \(\lambda_{\text{max}}\) is the largest eigenvalue of \(J'\)), whereas \(a(t)\) is a linearly increasing function of time, which drives the system through the bifurcation point, occurring approximately when \(a(t) = a_0\). Post-bifurcation, the energy landscape governing the evolution of particles approximately encodes the local minima of the Ising term, and thus the particles flow towards the low-energy solutions of the optimization problem, which then can be extracted by taking the sign \(s_i = \mathrm{sign}(q_i)\). In an effort to suppress the errors originating from relaxing the variables from binary to continuous, a modified version of SBM was introduced\cite{goto2021sbm} (which we use in this work), where the nonlinear term \(q_i^4\) was replaced by perfectly inelastic walls, located at \(|q_i| = 1\), and the Ising contribution to dynamics was discretized, i.e. term \(\sum_{j=1}^N J'_{ij}q_{j}\) in Eq.~\eqref{eq:sbm} got replaced by \(\sum_{j=1}^N J'_{ij}\mathrm{sign}(q_{j})\). Additionally, these modifications guarantee that the dynamics can be stopped when \(a(t) = a_0\), and the system will be located in a local minimum. 

On the computational side, differential equations \eqref{eq:sbm2}-\eqref{eq:sbm} are separable with respect to positions and momenta, and thus amenable to the numerically stable, symplectic Euler integration scheme, which can be efficiently implemented on GPUs. Moreover, the chaotic nature of the SBM equations imply sensitivity to initial conditions, and because each run is independent, many starting points can be evolved simultaneously, further increasing the efficiency and potential for massive parallelization.

\section{Conclusions}\label{sec::conclusions}

In this work, we have demonstrated the utility of a hybrid quantum-classical approach for a near-real-life AGVs scheduling problem that can be implemented in a real factory environment. While we have not yet demonstrated a quantum advantage, one of the hybrid quantum-classical approaches produced results close to those of CPLEX, which is widely recognized as the benchmark classical solver. As these hybrid approaches are expected to improve in the near future, we may be on the brink of, at least, achieving quantum advantage.

Larger problems involving hundreds of AGVs, anticipated with the development of Industry 4.0 (resulting in several thousand variables in our model), may remain beyond the capabilities of classical algorithms. However, future hybrid quantum-classical algorithms are expected to have the potential to handle these larger problems.

The CQM solver, even for large problems, provides a range of diverse solutions, which can be used as input for multi-case decision support systems or sophisticated stochastic scheduling approaches. In contrast, while CPLEX (with limited time) is faster and closer to optimality, it provides only one solution. Our experience with BQM highlights the importance of systematic penalty determination as discussed in recent studies~\cite{garcia2022exact}~\cite{quintero2023polyhedral}.

We recognize the significant methodological value in contributions that utilize open-source tools or develop custom hybrid classical-quantum workflows. To our knowledge, no production-quality general hybrid quantum-classical solvers that input native ILP are currently available. In contrast, for the QUBO implementation, we applied the open-source SBM solver, which outperforms the BQM solver in terms of solution quality but still fails to return feasible solutions (except for the smallest cases). Since the SBM solver is quantum-inspired and does not use quantum resources, our findings suggest that quantum annealing does not play a predominant role in the iterative hybrid process, aligning with the "Quantum Backstops" hypothesis. These observations highlight the need for more transparent hybrid quantum-classical or quantum-inspired solvers to effectively benchmark quantum (or quantum-inspired) computation. We believe it is important for the quantum computing scientific community to systematically work on these developments.

\section*{Data availability}
The code and the data used for generating the numerical results can be found in \url{https://github.com/iitis/AGV_quantum}

\section*{Acknowledgements} 
The research was supported by the Foundation for Polish Science (FNP) under grant number TEAM NET POIR.04.04.00-00-17C1/18-00 (BG, ZP, ŁP); National Science Centre, Poland under grant number 2022/47/B/ST6/02380 (KD), and under grant number 2020/38/E/ST3/00269 (TŚ), and by the Ministry of Culture and Innovation and the National Research, Development and Innovation Office within the Quantum Information National Laboratory of Hungary (Grant No. 2022-2.1.1-NL-2022-00004) (MK). S.D. acknowledges support from the John Templeton Foundation under Grant No. 62422. For the purpose of Open Access, the authors have applied a CC-BY public copyright license to any Author Accepted Manuscript (AAM) version arising from this submission.

\section*{Author contributions statement}
Z.P., B.G., S.D., K.D. - conceptualization, K.D., T.Ś., J.P., A.P. - preparing experiments, T.Ś., J.P., A.P. - running experiments,  K.D., Z.P., Ł.P., M.K., J.P., A.P., B.G., S.D. - data analysis, K.D., 
M.K, T.Ś., J.P. - manuscript writing, Z.P., Ł.P., B.G., S.D. - manuscript supervision. All authors reviewed the manuscript.

\appendix

\section{Details of one particular AGVs problem}\label{sec::appendix_details}

In this Appendix, we present a step-by-step solution of the $7$ AGVs and $7$ zones problem with topology presented in Fig~\ref{fig::network}. We have bi-directional lanes between zones $s_0$, $s_1$, $s_2$, $s_3$, $s_4$ and $s_5$ as well uni-directional lane between zones $s_5$ and $s_6$.
For this particular computation, the following parameters have been chosen:
$\tau^{\text{headway}} = 2$ (for all pairs of AGVs) $\tau^{\text{zone}} = 2$ (for all AGVs), passing times (for each AGV):  $\tau^{\text{pass}}(s_0, s_1) = \tau^{\text{pass}}(s_1, s_2) = \tau^{\text{pass}}(s_3, s_4) = 6$, $\tau^{\text{pass}}(s_2, s_3) = 0$, $\tau^{\text{pass}}(s_4, s_5) = \tau^{\text{pass}}(s_5, s_6) = 4$. 

Then, we assume that each AGV (denoted by $j$) is ready to enter its initial station (denoted by $s_{j,0}$) at $v_{in}(j, s_{j,0})$.
We also assume that AGVs have the following paths and initial conditions: \begin{equation}
    \begin{split}
     &\text{AGV}_0: \{s_0, s_1, s_2, s_3 \},    v_{in}(j_0, s_0) = 0 \ \ \ \ \ \ \ \ \
     \text{AGV}_1: \{s_0, s_1, s_2\}, \ \ \ \  \ \ \ \ \  v_{in}(j_1, s_0) = 0 \\
     &\text{AGV}_2: \{s_4, s_3, s_2, s_1\},  v_{in}(j_2, s_4) = 8 \ \ \ \ \ \ \ \ \
     \text{AGV}_3: \{s_4, s_3, s_2, s_1, s_0\},  v_{in}(j_3, s_4) = 9 \\
     &\text{AGV}_4: \{s_2, s_3  \} \ \ \ \  \ \ \ \ \ \  \ v_{in}(j_4, s_2) = 15 \ \ \ \ \ \ \ \
     \text{AGV}_5: \{s_6, s_5, s_4, s_3  \} \ \  \ \ \  \ v_{in}(j_5, s_6) = 0 \\
     &\text{AGV}_6: \{s_5, s_6  \} \ \ \ \  \ \ \ \ \ \  \ v_{in}(j_6, s_5) = 0 \\
    \end{split}
\end{equation}
In practice, the above will be done by step $2$ of Algorithm in Section~\ref{sec::algorithm}.
In the objective function in Eq.~\eqref{eq::objective} for this computational example each AGV is assigned an equal weight of $w_j = 1$. Further, we set $d_{\max} = 40$.

\begin{figure}[ht]
\centering
    \includegraphics[width = 0.31\linewidth]{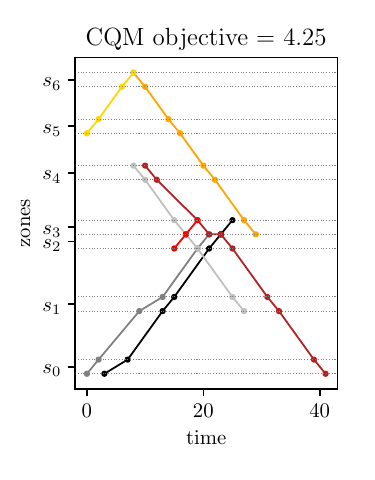}
    \includegraphics[width = 0.31\linewidth]{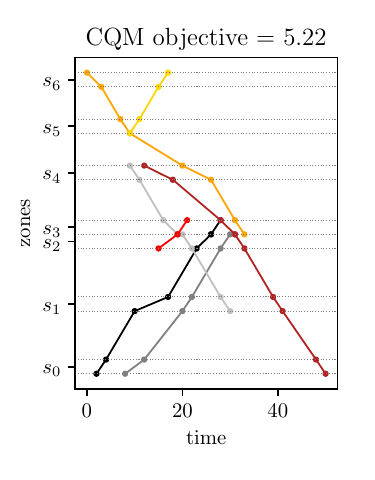}
    \includegraphics[width = 0.31\linewidth]{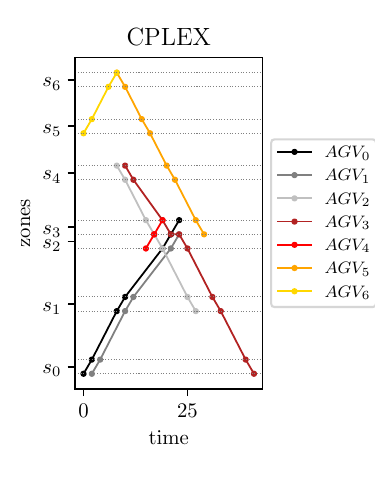}
    \caption{Solution in the form of a space-time diagram, two suboptimal CQM solutions and an optimal one obtained with both CQM and CPLEX. The suboptimal solutions can also be useful for the decision support system.}\label{fig::t_diagram}
\end{figure}

The most important step of the Algorithm in Section~\ref{sec::algorithm} is step $5$: solving the optimization problem. The  CQM hybrid solver yields many solutions at a time.
The best solution coincides with the optimal one also provided by  CPLEX. However other solutions can also be of use for the decision support system. 
The solutions are presented in the form of a simplified time-space diagram (in which known points are connected with lines) in Fig.~\ref{fig::t_diagram}.

\end{document}